\documentclass{article}  
\usepackage{jamaica04}
\usepackage{graphicx}
\usepackage{epsf}
\usepackage{amssymb}
\frompage{000} \topage{000}                                              
\def\rightmark{Study on chemical equilibrium in nucleus-nucleus collisions...}
\def\leftmark{J. Manninen {\it et al.}}

\def\Q1{{\bf Q}_1}

\def\qj{{\bf q}_j}

\def\muv{\mbox{\boldmath$\mu$}}
\def\muvs{\mbox{\boldmath${\scriptstyle{\mu}}$}}
\def\d{{\rm d}}

\def\e{{\rm e}}

\def\gs{\gamma_S}
\def\gss{\gamma_s}
\def\gq{\gamma_q}
\def\ls{\lambda_S}
\def\vexp{$VT^3 \, \exp[-0.7 \, {\rm GeV}/T]$}

\def\ssb{\langle {\rm s}\bar{\rm s}\rangle}
\def\uub{\langle {\rm u}\bar{\rm u}\rangle}
\def\ddb{\langle {\rm d}\bar{\rm d}\rangle}

\def\kpi{$\langle {\rm K}^+ \rangle / \langle \pi^+ \rangle$ }

\title{Study on chemical equilibrium in nucleus-nucleus collisions at relativistic energies}  

\abstract{We present a detailed study of chemical freeze-out in nucleus-nucleus collisions 
at beam energies of 11.6, 30, 40, 80 and 158$A$ GeV. By analyzing hadronic 
multiplicities within the statistical hadronization approach, we have studied
the chemical equilibration of the system as a function of center of mass energy and of
the parameters of the source. Additionally, we have tested and compared different versions of 
the statistical model, with special emphasis on possible 
explanations of the observed strangeness hadronic phase space under-saturation. }

\authors{
{ J.Manninen$^1$, F. Becattini$^2$, A. Ker\"anen$^{1,2}$, M. Ga\'zdzicki$^{3,4}$, 
and R. Stock$^{3,5}$}\\[2.812mm]
{\normalsize
\hspace*{-8pt}$^1$ University of Oulu, Oulu, Finland\\[0.2ex]
\hspace*{-8pt}$^2$ Department of Physics, Universit\`a di 
Firenze and INFN Sezione di Firenze, \\ 
Firenze, Italy\\[0.2ex] 
\hspace*{-8pt}$^3$ Institut f\"ur  Kernphysik, Universit\"at Frankfurt, Frankfurt, Germany\\[0.2ex] 
\hspace*{-8pt}$^4$ \'Swi\c{e}tokrzyska Academy, Kielce, Poland\\[0.2ex]
\hspace*{-8pt}$^5$ CERN, Geneva, Switzerland
}}


\begin{document}
\addtolength{\headsep}{-10pt}
\addtolength{\footskip}{-10pt}
\addtolength{\textheight}{20pt}
\vspace{-0.7cm} 
\maketitle

\setcounter{page}{1}

\vspace{-0.5cm}

\section{The Statistical Hadronization Model}

\vspace{-0.3cm}
One of the main results of the study of high energy A-A collisions is a surprising 
success of the statistical-thermal models in reproducing essential features of 
particle production (see for example~\cite{beca2,becapt}).
In this paper we study nucleus-nucleus collision
within the statistical model in the energy range that is believed~\cite{Gaz1998vd} to 
cover the threshold for creation of Quark-Gluon Plasma (QGP) in the early stage 
of Pb-Pb collisions.

The main idea of the SHM is that hadrons are emitted from regions at statistical 
equilibrium. No hypothesis is made about how 
statistical equilibrium is achieved; this can be a direct consequence of the 
hadronization process. In a single collision event, there might 
be several clusters with different collective momenta, different
overall charges and volumes. However, Lorentz-invariant quantities like particle 
multiplicities are independent of clusters momenta.
 
If the system size is sufficiently large~\cite{kerabeca}, 
the analysis can be done in grand-canonical 
ensemble in which the mean {\em primary} multiplicity of the 
$j^{\rm th}$ hadron with mass $m_j$ and spin $J_j$ reads: 
\vspace{-0.3cm}
\begin{equation}\label{mean}
\left< n_j \right> = \frac{(2J_j+1) V }{(2\pi)^3} \int \d^3 {\rm p} \; 
\left[ \e^{\sqrt{{\rm p}^2+m_j^2}/T+\muvs\cdot\qj/T} \pm 1 \right]^{-1}
\end{equation}
where $T$ is the temperature, $V$ the scaling volume, $\qj = (Q_j,B_j,S_j)$ is a 
vector having as components the electric charge, baryon number and strangeness 
of the hadron and $\muv = (\mu_Q,\mu_B,\mu_S)$ is a vector of the corresponding 
chemical potentials; the upper sign applies to fermions, the lower to bosons.

In order to correctly reproduce the data, it is also necessary to introduce at 
least one non-equilibrium parameter suppressing hadrons containing valence strange 
quarks, $\gamma_S \neq 1$ \cite{gammas}. With this supplementary parameter, hadron 
multiplicity is as in Eq.~(\ref{mean}) with the replacement:
$\exp[\muv\cdot\qj/T] \rightarrow \exp[\muv\cdot\qj/T] \gamma_S^{n_s}$
where $n_s$ stands for the number of valence strange quarks {\em and} anti-quarks 
in the hadron $j$.  

Finally, the overall  
multiplicity to be compared with the data, is calculated as the sum of primary 
multiplicity (\ref{mean}) and the contribution from the decay of heavier hadrons:
$\langle n_j \rangle  = \langle n_i \rangle^{\mathrm{primary}} + 
\sum_k \mathrm{Br}(k\rightarrow j) \langle n_k \rangle$,
where the branching ratios are taken from the latest issue of the Review of Particle 
Physics \cite{pdg}.

\vspace{-0.4cm}

\section{Experimental Data Set And Analysis Results}

\vspace{-0.3cm}

The bulk of the experimental data consists of measurements made by NA49 
collaboration in central Pb-Pb collisions at beam momenta of 30, 40, 80 and 
158$A$ GeV~\cite{PbPb,Af2002mx}. As far as AGS data at 11.6$A$ 
GeV~\cite{centrags,lamb896,lamb891,beca01} 
is concerned, 
we have used both multiplicities measured by the experiments and extrapolations of measured 
rapidity distributions (for details, see~\cite{paper}). 

The analysis has been carried out by looking for the minima of the \\ 
$\chi^2 = \sum_i \frac{(n_i^{\rm exp} - n_i^{\rm theo})^2}{\sigma_i^2}$.
The fitted parameters within the main scheme SHM($\gs$) are shown in table~\ref{parameters}.
The observed differences in the fit parameters between two independent analyses A and B 
are of the order of the fit errors. 

A major result of these fits is that $\gs$ is significantly smaller than 1
in almost all cases, that is strangeness seems to be under-saturated with respect 
to a completely chemically equilibrated hadron gas. This confirms previous 
findings \cite{beca01,bgs,cley}.

\vspace{-0.3cm}
\subsection{{\bf Strangeness correlation volume}}
\vspace{-0.2cm}
To account for the observed under-saturation of strangeness, a picture has been
put forward in which strangeness is supposed to be exactly vanishing over 
distances less than those implied by the overall volume $V$ \cite{redl}, i.e. all 
clusters or fireballs emerge with $S=0$ and they are not allowed to share non-vanishing 
net strangeness. 

Assuming, for sake of simplicity, that all clusters have the same 
typical volume $V_c$ and that treating baryon number and electric charge (but not strangeness) 
grand-canonically, gives a good description of the system, 
one can perform the analysis in {\em strangeness canonical ensemble} (for details, see~\cite{paper}).

If $V_c$ is sufficiently small, the multiplicities of strange hadrons turn out to
be significantly suppressed with respect to the corresponding grand-canonical
ones due to an effect called {\em canonical suppression}. 

We have fitted 
the data sample of full phase
space multiplicities in Pb-Pb collisions at 158$A$ GeV fixing $\gs=1$.
The quality of the fit is worse ($\chi^2$/dof = 37.2/9) with respect to the SHM($\gs$) model, 
whilst thermal parameters $T$=157.9 MeV and $\mu_B$=261.5 MeV are compatible with the
main version of the statistical model. 
Our result suggests that, for the local strangeness correlation to be an 
effective mechanism, the cluster volume should be of the order of 2.5\% of the 
overall volume. 
 \vspace{-0.4cm}
\subsection{{\bf Superposition of NN collisions with a equilibrated fireball}}
\vspace{-0.25cm}
In this picture, henceforth referred to as SHM(TC), the observed hadron production 
is approximately the superposition of two components: one originated from a large 
fireball at complete chemical equilibrium at freeze-out, with $\gs=1$, and another 
component from single nucleon-nucleon collisions. 
Since it is known that in NN collisions strangeness is strongly suppressed~\cite{beca2},
the idea is to ascribe the observed under-saturation of strangeness in heavy
ion collisions to the NN component.

With the simplifying assumption of disregarding 
subsequent inelastic collisions of particles produced in those primary NN collisions, 
the overall hadron multiplicity can be written then as 
$ \langle n_j \rangle = \langle N_c \rangle \langle n_j \rangle_{NN} +
 \langle n_j \rangle_V,$
where $\langle n_j \rangle_{NN}$ is the average multiplicity of the $j^{\rm th}$
hadron in a single NN collision, $\langle N_c \rangle$ is the mean number of
single NN collisions and $\langle n_j 
\rangle_V$ is the average multiplicity of hadrons emitted from the equilibrated
fireball.

To calculate $\langle n_j \rangle_{NN}$ we have used the statistical model and
fitted pp full phase space multiplicities measured at the same beam energy by the same NA49 experiment. 
For np and nn collisions, the 
parameters of the statistical model determined in pp are retained and the 
initial quantum numbers are changed accordingly. Theoretical multiplicities have 
been calculated in the canonical ensemble, which is described in detail in 
ref.~\cite{becapt}. 

We have fitted
$T$, $V$, $\mu_B$ of the central fireball and $\langle N_c \rangle$ by using 
NA49 data in Pb-Pb collisions at 158$A$ GeV. 
The fit quality, as well as the obtained values of $T$ and $\mu_B$, are comparable 
to the main fit within the SHM($\gs$) model. The predicted number of
"single" NN collisions is about 50 with a 16\% uncertainty. 

\vspace{-0.4cm}  
\subsection{{\bf Non-equilibrium of hadrons with light quarks}}
\vspace{-0.25cm}

In this model~\cite{rafe} two non-equilibrium parameters are introduced for the different types 
of quarks, $\gq$ for u, d quarks and $\gss$ for strange quarks.
By defining:
$\gs = \frac{\gamma_s}{\gq}$ and $\tilde V = V \gq^2$,
the Boltzmann limit of average multiplicity reads:
\vspace{-0.3cm}
\begin{equation}\label{meangq2}
  \langle n_j \rangle = \frac{(2J_j + 1) {\tilde V}}{(2\pi)^3} 
  \gs^{n_s} \gq^{|B_j|} \int \d^3 {\rm p} \; \exp[-\sqrt{{\rm p}^2+m_j^2}/T
  + \muv \cdot \qj/T]
\end{equation}
where $B_j$ is the baryon number, as long as mesons have two and baryons have three 
valence quarks.
The parameter 
$\gq$ has a definite physical bound for bosons which can be obtained by requiring 
the convergence of the series $\sum_{N=0}^\infty (\gq^{n_q N}) \exp(-N \epsilon/T 
+ N \muv \cdot \qj/T)$ for any value of the energy. If  $\gq$ reaches its bounding value\\
$\gq = \exp(m_{\pi^0}/2T) \simeq 1.5$ for $T \simeq 160$ MeV, a Bose 
condensation of particles in the lowest momentum state sets in.
 
It is seen that the absolute $\chi^2$ minimum falls in the region of pion condensation, 
at $\gq \simeq 1.62$, with $\chi^2 \simeq 13$ and $T \simeq
140$ MeV. This finding is in agreement with what is found in ref.~\cite{rafe}. 
However, there is also a local minimum at the lower edge $\gq = 0.6$, with a temperature
of 187 MeV, which is only one unit of $\chi^2$ higher than the absolute minimum. 
This indicates that the absolute minimum could be rather unstable against variations
of the input data and this is in fact what we find by varying down the pion multiplicities
by only 1 $\sigma$. For this case, the absolute minimum of $\chi^2$ now lies at $\gq = 0.6$ 
instead of at the pion condensation point.  
  
In view of the instability of the fit, and of the small {\em relative} $\chi^2$ 
improvement in comparison with the main fit, we conclude that there is so far 
no evidence for the need of this further non-equilibrium parameter. 

\vspace{-0.5cm}
 
\section{Energy Dependence} 

\vspace{-0.3cm}

\begin{center}
\begin{figure}[!ht]
$\begin{array}{c@{\hspace{1in}}c}
\epsfxsize=2.6in
\epsffile{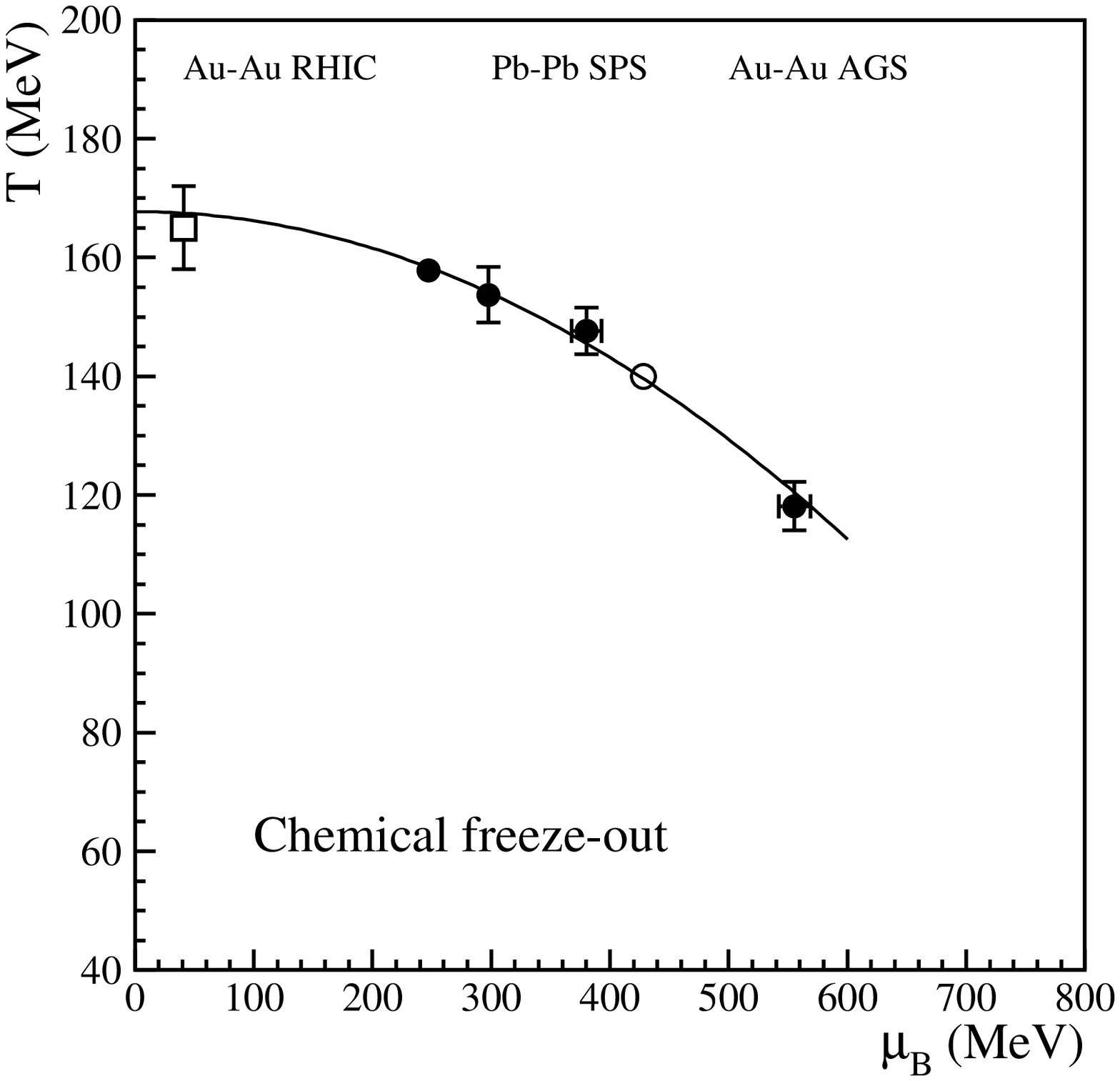} & \hspace*{-3cm}
	\epsfxsize=2.6in
	\epsffile{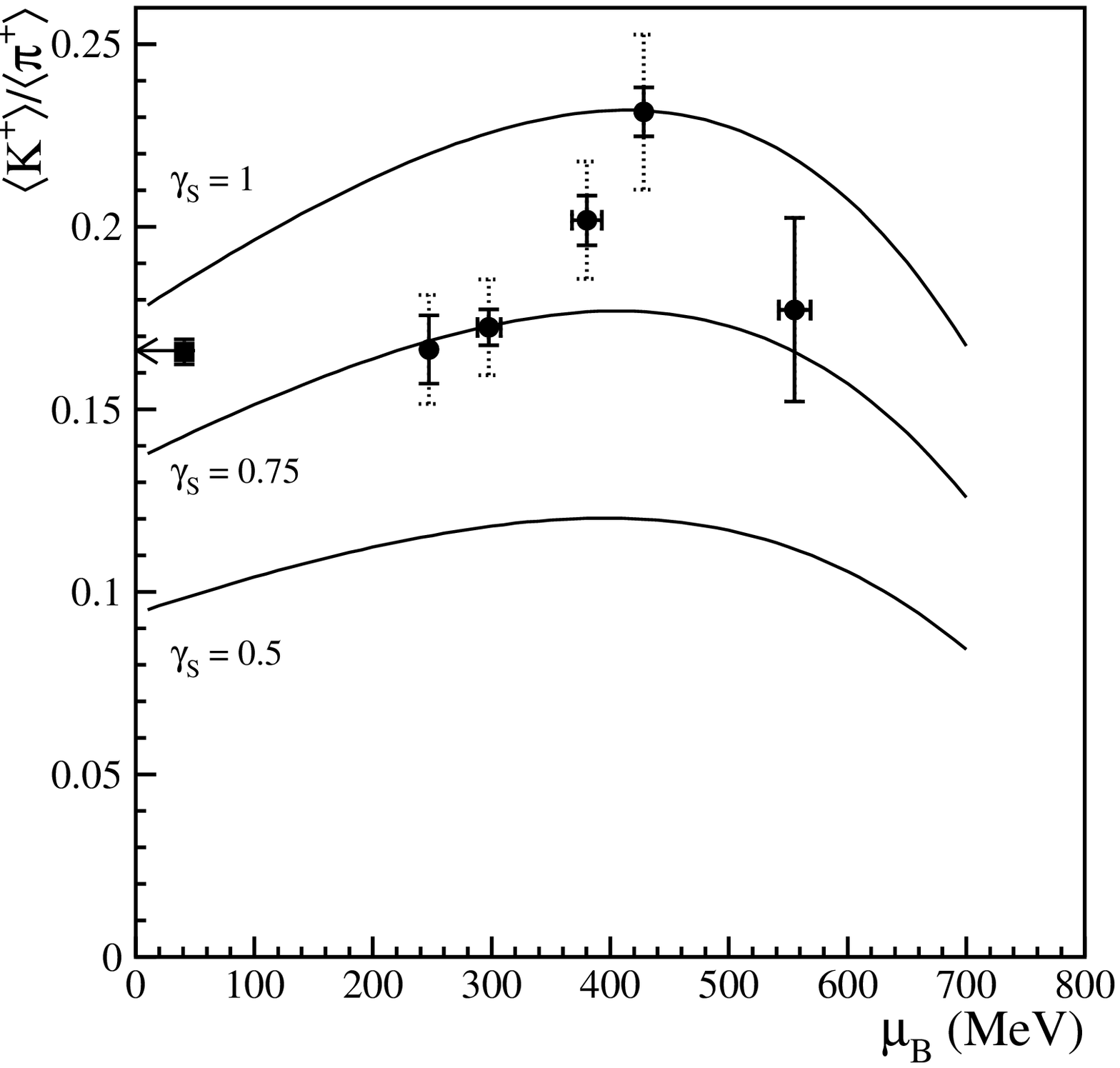} \\ 
\end{array}$
\vspace{-0.6cm}
\caption{LEFT: Chemical freeze-out points in the $\mu_B-T$ plane in various heavy ion 
collisions. The full round dots refer to Au-Au at 11.6 and Pb-Pb collisions 
at 40, 80, 158$A$ GeV obtained in the analysis A, whilst the hollow square dot 
has been obtained in ref.~\cite{flork} by using particle ratios measured at 
midrapidity in Au-Au collisions at $\sqrt s_{NN} = 130$ GeV.
The hollow round dot without error bars refers to Pb-Pb collisions at 
30$A$ GeV and has been obtained by forcing $T$ and $\mu_B$ to lie on the
parabola fitted to the full round dots.\newline
RIGHT: Measured \kpi ratio as a function of the fitted baryon-chemical 
potential. The full square dot is a preliminary full phase space measurement
in Au-Au collisions at $\sqrt s_{NN} = 200$ GeV \cite{jordre}. For the SPS
energy points the statistical errors are indicated with solid lines, while the
contribution of the common systematic error is shown as a dotted line. Also 
shown the theoretical values for a hadron gas along the fitted chemical 
freeze-out curve (left), for different values of $\gs$.}\label{tmukpi}
\vspace{-0.3cm}
$\begin{array}{c@{\hspace{1in}}c}
\epsfxsize=2.6in
\epsffile{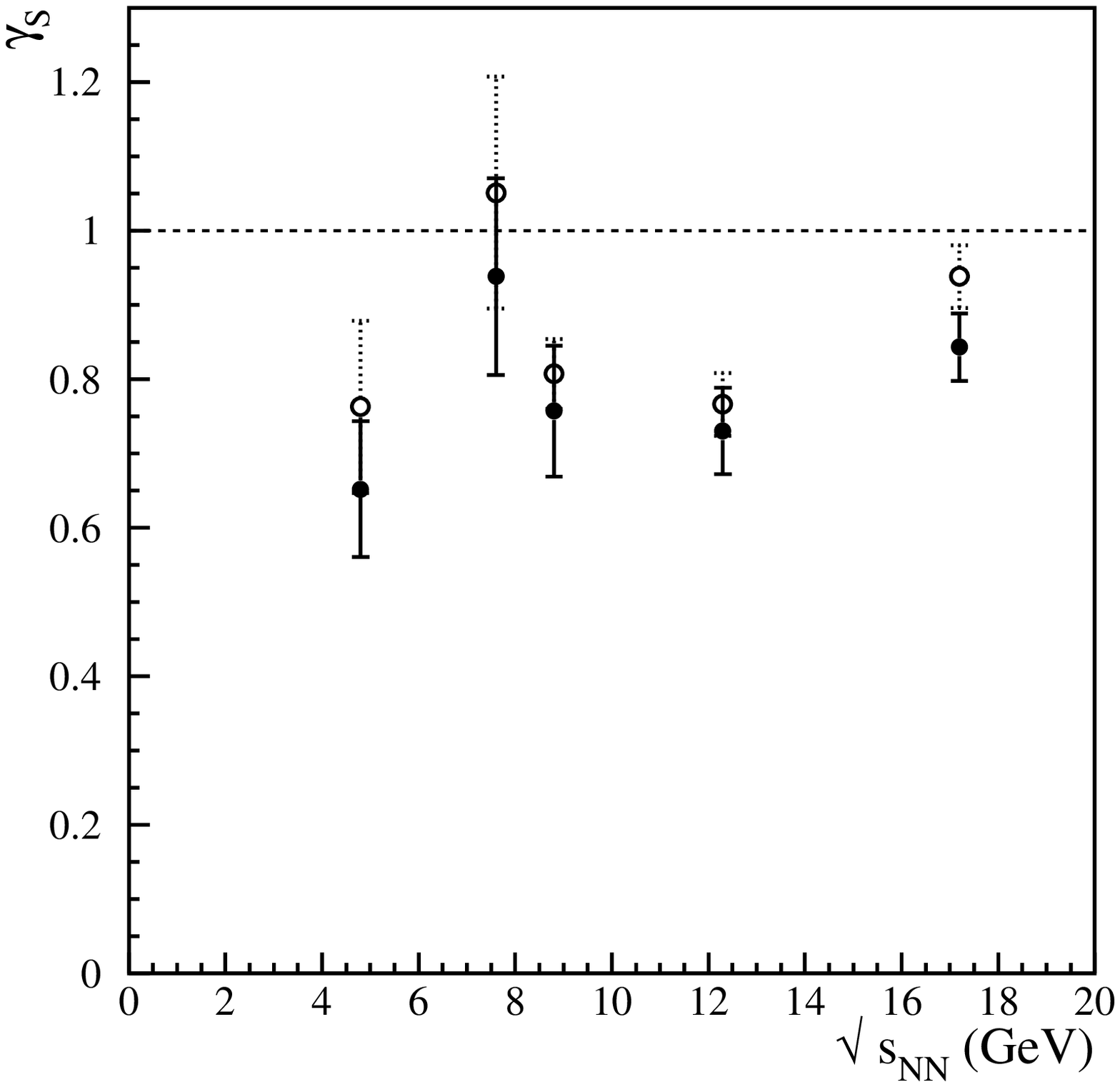} & \hspace*{-3cm}
	\epsfxsize=2.6in
	\epsffile{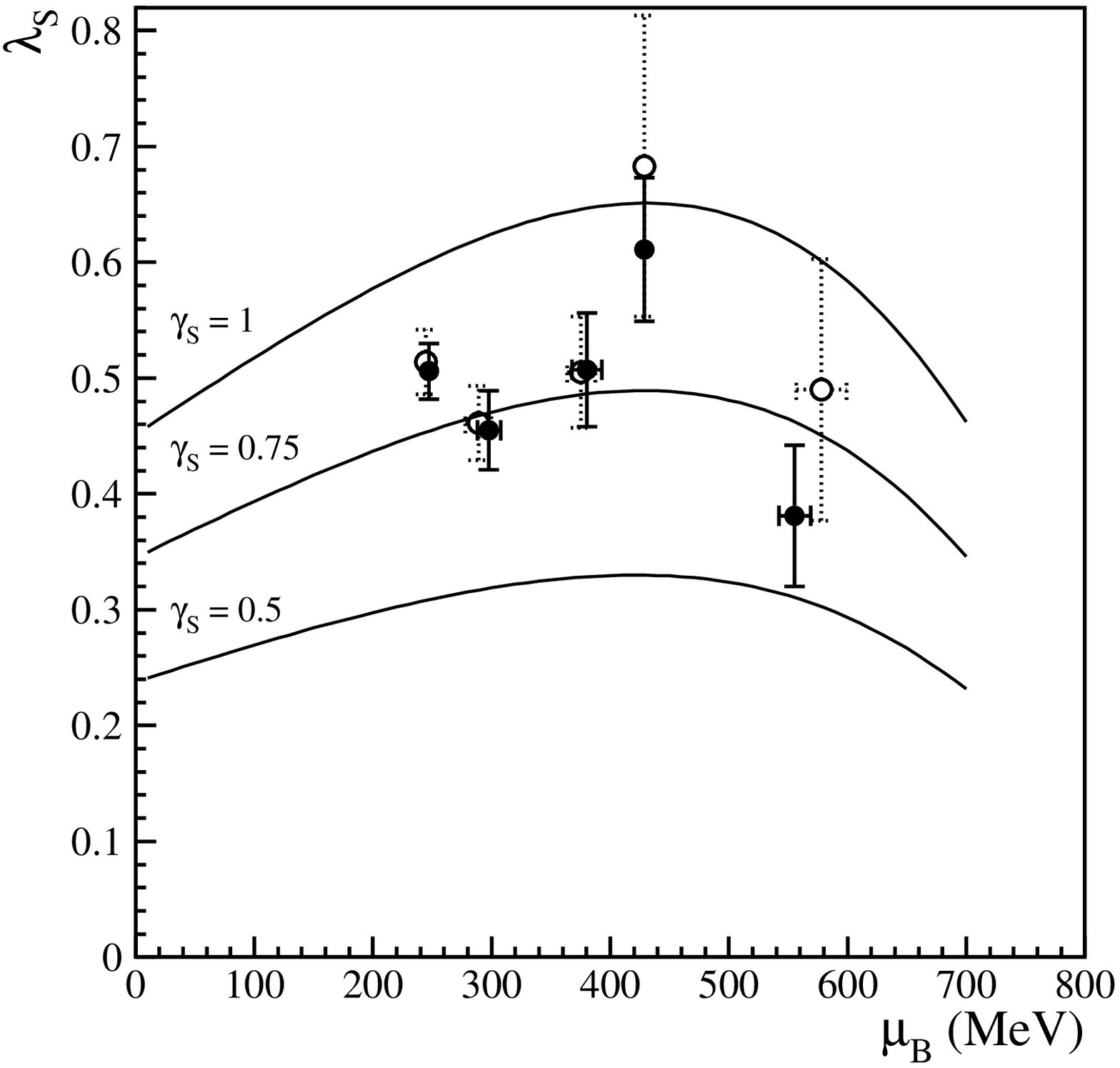} \\ 
\end{array}$
\vspace{-0.7cm}
\caption{LEFT: Strangeness non-equilibrium parameter $\gs$ as a function of the nucleon-nucleon
centre-of-mass energy. Full dots refer to fit A, hollow dots fit B.\newline
RIGHT: $\ls$ estimated as a 
function of the fitted baryon-chemical potential. Also shown the theoretical 
values for a hadron gas along the fitted chemical freeze-out curve shown in 
fig.~\ref{tmukpi}, for different values of $\gs$. \label{gsls}}
\end{figure}
\end{center}

The chemical freeze-out points in the $\mu_B-T$ plane are shown in fig.~\ref{tmukpi}.
The four points
at beam energies of 11.6, 40, 80 and 158$A$ GeV have been fitted with a parabola:
 $T = 0.167 - 0.153 \mu_B^2$,
where $T$ and $\mu_B$ are in GeV. 

A possible indication of deconfinement phase transition  in Pb-Pb collisions at the low  
SPS energies was reported on the basis of the observed energy dependence of 
several observables \cite{anomalies}. Particularly, the \kpi ratio shows a peaked 
maximum at about 30$A$ GeV. One may expect that this anomaly should be reflected 
in the energy dependence of $\gs$ parameter fitted within SHM($\gs$) scheme.
This dependence is plotted in fig.~\ref{gsls} and in fact a maximum shows up at 
30$A$ GeV. 

The anomalous increase of relative strangeness production at 30$A$ GeV 
can be seen also in the Wroblewski variable $\ls = 2 \ssb/(\uub+\ddb)$, the estimated 
ratio of newly produced strange quarks to u, d quarks at primary hadron level, 
shown in fig.~\ref{gsls}

In order to further study strangeness production features, we have also compared the
the measured \kpi ratio  with the theoretical values in a hadron gas along 
the interpolated freeze-out curve for different values of $\gs$ (see fig.~\ref{tmukpi}). 
The calculated dependence of \kpi on 
$\mu_B$ is non-monotonic with a broad maximum at $\mu_B \simeq 400$ MeV (i.e. 
$E_{beam} \simeq 30A$ GeV). Taking into account that systematic
errors at different energies in Pb-Pb collisions are fully correlated, we can
conclude that the data points seem not to follow the constant $\gs$ lines.
\begin{center}
\begin{table}[!hb]
\hspace{-1cm}
\vspace{-0.2cm}
\begin{footnotesize}
\vspace{0.3cm}
\begin{tabular}{|c|c|c|c|c|c|}
\hline
 Parameters     & Main analysis A       & Main analysis B   &       &   Main analysis A       & Main analysis B   \\
\hline
\hline               
 & \multicolumn{2}{|c|}{Au-Au 11.6$A$ Gev} & & \multicolumn{2}{|c|}{Pb-Pb 30$A$ GeV} \\
\hline               
$T$ (MeV)       & 118.1$\pm$3.5 (4.1)     & 119.1$\pm$4.0 (5.4) &  &  139.5                   & 140.3  \\
$\mu_B$ (MeV)   & 555$\pm$12 (13)         & 578$\pm$15 (21)     &     &  428.6                   & 428.7 \\ 
$\gs$           & 0.652$\pm$0.069 (0.079) & 0.763$\pm$0.086 (0.12) &  &  0.938$\pm$0.078 (0.13)  & 1.051$\pm$0.103 (0.16)  \\
V'           & 1.94$\pm$0.21 (0.24)    & 1.487$\pm$0.18 (0.25)   &    & 6.03$\pm$0.50 (0.85)    & 5.273$\pm$0.526 (0.80)  \\ 
\hline
$\chi^2$/dof    & 4.0/3                   & 5.5/3                &    &   5.75/2                  & 4.6/2  \\
\hline
\hline               
 & \multicolumn{2}{|c|}{Pb-Pb 40$A$ GeV} & & \multicolumn{2}{|c|}{Pb-Pb 80$A$ GeV} \\
\hline
$T$ (MeV)       & 147.6$\pm$2.1 (4.0)     & 145.5$\pm$1.9 (3.5) &     & 153.7$\pm$2.8 (4.7)      & 151.9$\pm$3.4 (5.4)  \\
$\mu_B$ (MeV)   & 380.3$\pm$6.5 (13)      & 375.4$\pm$6.4 (12)   &    &  297.7$\pm$5.9 (9.8)      & 288.9$\pm$6.8 (11) \\
$\gs$           & 0.757$\pm$0.024 (0.046) & 0.807$\pm$0.025 (0.047) &  &  0.730$\pm$0.021 (0.035)  & 0.766$\pm$0.026 (0.042)\\
V'           & 8.99$\pm$0.37 (0.71)    & 8.02$\pm$0.34 (0.63)     &   & 15.38$\pm$0.61 (1.0)     & 14.12$\pm$0.65 (1.1) \\
\hline
$\chi^2$/dof    & 14.7/4                   & 13.6/4               &   &  11.0/4                   & 10.4/4       \\
\hline
\hline               
 & \multicolumn{2}{|c|}{Pb-Pb 158$A$ GeV} & & & \\
\hline                
$T$ (MeV)       & 157.8$\pm$1.4 (1.9)      & 154.8$\pm$1.4 (2.1)   &   & & \\
$\mu_B$ (MeV)   & 247.3$\pm$5.2 (7.2)      & 244.5$\pm$5.0 (7.8)    &  & &\\
$\gs$           & 0.843$\pm$0.024 (0.033)  & 0.938$\pm$0.027 (0.042) & & & \\
V'           & 21.13$\pm$0.80 (1.1)     & 18.46$\pm$0.69 (1.1)     & & & \\
\hline
$\chi^2$/dof    & 16.9/9                   & 21.6/9                   & & & \\
\hline               
\end{tabular}
\end{footnotesize}
\vspace{-0.5cm}
\caption{Summary of fitted parameters (V'=\vexp) at AGS and 
SPS energies in the framework of the SHM($\gs$) model.
The re-scaled errors (see~\cite{pdg,paper}) are quoted within brackets. For Pb-Pb at 30 $A$ GeV 
data, we have constrained $T$ and $\mu_B$ to lie on the fitted chemical 
freeze-out curve (see fig.~\ref{tmukpi})\label{parameters}}
\end{table}
\end{center}

\vspace{-1.0cm}
\section{Summary and Conclusions}

\vspace{-0.3cm}

It is found that the main version of the statistical model (with $\gs$), fits all the 
data analyzed in this paper.
We have tested a model (SHM(TC)) 
which can fit the data at 158$A$ GeV very well if the number of independent NN collisions is 
around 50 with a sizeable uncertainty. 
A model in which strangeness is assumed to vanish locally \cite{redl} yields
a worse fit to the data with respect to SHM($\gs$) and SHM(TC).    
Moreover, we have found that the present set of available data does not allow to 
establish whether
a further non-equilibrium parameter ($\gq$) is indeed needed to account for the 
observed hadron production pattern.

The evolution of the freeze-out temperature and baryon-chemical potential
is found to be smooth in the AGS-SPS-RHIC energy range. \kpi ratio, Wroblewski 
factor $\ls$ as well as $\gs$ parameter calculated within the statistical model 
suggests that there might be a 
peak in relative strangeness production at about 30$A$ GeV of beam momenta.

\clearpage  


\vspace{-0.6cm}

\begin{thebibliography}{99} 

\vspace{-0.4cm}

\bibitem{beca2}
 F.~Becattini, U.~Heinz, Z. Phys. C {\bf 76} (1997) 269.

\bibitem{becapt}
 F.~Becattini, G.~Passaleva, Eur. Phys. J. C {\bf 23} (2002) 551.

\bibitem{Gaz1998vd} 
M.~Gazdzicki and M.~I.~Gorenstein, Acta Phys.\ Polon.\ B {\bf 30} (1999) 2705.

\bibitem{kerabeca} 
 A. Ker\"anen and F. Becattini, Phys. Rev. C {\bf 65} (2002) 044901.

\bibitem{gammas} 
 P. Koch, B. M\"uller and J. Rafelski, Phys. Rep. {\bf 142} (1986) 167.

\bibitem{pdg} 
 K. Hagiwara {\it et al.}, Phys. Rev. D {\bf 66} (2002) 010001-1.
  
\bibitem{PbPb}
C.~Alt {\it et al.}, nucl-ex/0305017, S.~V.~Afanasiev {\it et al.}, 
Nucl.\ Phys.\ A {\bf 715} (2003) 161 and 453,  V.~Friese, NA49 Coll., Nucl.\ Phys.\ A 
{\bf 698} (2002) 487.

\bibitem{Af2002mx} 
S.~V.~Afanasiev {\it et al.}, NA49 Coll., Phys.\ Lett.\ B {\bf 491} (2000) 59,
Phys.\ Lett.\ B {\bf 538} (2002) 275, Phys. Rev. C {\bf 66} (2002) 054902.

\bibitem{centrags}
 L.~Ahle {\it et al.}, E-802 Coll., Phys.\ Rev.\ C {\bf 60} (1999) 044904 and 064901.

\bibitem{lamb896}
 S.~Albergo {\it et al.}, Phys.\ Rev.\ Lett.\ {\bf 88} (2002) 062301.

\bibitem{lamb891}
 S.~Ahmad {\it et al.}, Phys.\ Lett.\ B {\bf 382} (1996) 35.

\bibitem{beca01}
F. Becattini {\it et al.}, Phys. Rev. C {\bf 64} (2001) 024901.

\bibitem{bgs}
 F. Becattini, M. Ga\'zdzicki, J. Sollfrank, Eur. Phys. J. C {\bf 5} (1998) 143.

\bibitem{cley}
 J~Cleymans, B.~Kaempfer and S.~Wheaton, Phys.\ Rev.\ C {\bf 65}, 027901 (2002).

\bibitem{paper}
F. Becattini {\it et al.}, Phys. Rev. C {\bf 69} (2004) 024905.

\bibitem{redl}
 S.~Hamieh, K.~Redlich and A.~Tounsi, Phys.\ Lett.\ B {\bf 486} (2000) 61.

\bibitem{rafe} 
 J. Letessier and J. Rafelski, Phys. Rev. C {\bf 59} (1999) 947.

\bibitem{anomalies}
 V. Friese et al., NA49 Coll., nucl-ex/0305017; M.~Gazdzicki, hep-ph/0305176.

\bibitem{flork} 
 W.~Florkowski {\it et al.}, Acta Phys.\ Polon.\ B {\bf 33}, 
 (2002) 761.

\bibitem{jordre} 
 J. I. Jordre, talk given at EPS2003 Conference, Aachen (Germany) July 2003.

\end{thebibliography}
\end{document}